\documentclass[12 pt]{amsart}
 \theoremstyle{definition}
 
 \theoremstyle{remark}

 \numberwithin{equation}{section}

\begin{document}
\openup .5 \jot
\title[Invertible quantum measurement maps]{ A geometric characterization of \\
invertible quantum measurement maps}

\author{Kan He}\address[Kan He]{Faculty of Mathematics, Institute of Mathematics,
Taiyuan University of Technology, Taiyuan,
 030024, P. R. China} \email[K. He]{kanhemath@yahoo.com.cn}
\author{Jin-Chuan Hou}\address[Jin-Chuan Hou]{Faculty of Mathematics, Institute of Mathematics, Taiyuan University of Technology, Taiyuan,
 030024, P. R. China} \email[J. Hou]{jinchuanhou@yahoo.com.cn}
\author{Chi-Kwong Li}\address[Chi-Kwong Li]{Department of Mathematics, College of William  Mary, Williamsburg,
VA 23187-8795, USA; Faculty of Mathematics, Institute of
Mathematics, Taiyuan University of Technology, Taiyuan,
 030024, P. R. China} \email[C.K. Li]{ckli@math.wm.edu}

\thanks{ This work is partially supported by National Natural Science Foundation of
China (11171249,  11271217, 11201329), a grant to International
Cooperating Research from Shanxi (2011081039). Li was also supported by
a USA NSF grant and a HK RCG grant.} \maketitle
\begin{abstract}

A geometric characterization is given for invertible quantum
measurement maps. Denote by ${\mathcal S}(H)$ the convex set of all
states (i.e., trace-1 positive operators) on Hilbert space $H$ with
dim$H\leq \infty$, and $[\rho_1, \rho_2]$ the line segment joining
two elements $\rho_1, \rho_2$ in ${\mathcal S}(H)$. It is shown that
a bijective map $\phi:{\mathcal S}(H) \rightarrow {\mathcal S}(H)$
satisfies $\phi([\rho_1, \rho_2]) \subseteq
[\phi(\rho_1),\phi(\rho_2)]$ for any $\rho_1, \rho_2 \in {\mathcal
S}$ if and only if $\phi$ has one of the following forms
$$\rho \mapsto \frac{M\rho M^*}{{\rm tr}(M\rho M^*)}\quad
\hbox{ or } \quad \rho \mapsto \frac{M\rho^T M^*}{{\rm tr}(M\rho^T
M^*)},$$ where $M$ is an invertible bounded linear operator and
$\rho^T$ is the transpose of $\rho$ with respect to an arbitrarily
fixed orthonormal basis.
\end{abstract}

\section{Introduction and the main result}

In the mathematical framework of the theory of quantum information,
a state is a positive operator of trace 1 acting on a complex
Hilbert space $H$. Denote by ${\mathcal S}(H)$ the set of all states
on $H$, that is, of all positive operators with trace 1. It is clear
that ${\mathcal S}(H)$ is a closed convex subset of ${\mathcal
T}(H)$, the Banach space of all trace-class operators on $H$ endowed
with the trace-norm $\|\cdot \|_{\rm Tr}$. In quantum information
science and quantum computing, it is important to understand,
characterize, and construct different classes of maps on states. For
instance, all quantum channels and quantum operations are completely
positive linear maps; in quantum error correction, one has to
construct the recovery map for a given channel; to study the
entanglement of states, one constructs entanglement witnesses, which
are special types of positive maps; see \cite{NC}. In this
connection, it is helpful to know the characterizations of maps
leaving invariant some important subsets or quantum properties.
Such questions have attracted the attention of many
researchers; for example, see \cite{A,BZ,F,Hou,m,m1,m2}.

In this paper, we characterize
invertible maps $\phi:{\mathcal S}(H)\rightarrow{\mathcal S}(H)$ that satisfies
$$\phi([\rho_1, \rho_2]) \subseteq [\phi(\rho_1), \phi(\rho_2)] \quad \hbox{ for any }
\rho_1, \rho_2 \in {\mathcal S}(H),$$ where $[\rho_1, \rho_2] = \{t
\rho_1 + (1-t) \rho_2: t \in [0,1]\}$ denotes the closed line
segment joining two states $\rho_1, \rho_2$. In other words, we
characterize maps on states such that for any
$\rho_1,\rho_2\in{\mathcal S}(H)$ and $0\leq t\leq 1$, there is some
$s$ with $0\leq s\leq 1$ such that
$$\phi(t\rho_1+(1-t)\rho_2)=s\phi(\rho_1)+(1-s)\phi(\rho_2).$$
This question is motivated by the study of affine
isomorphisms on ${\mathcal S}(H)$; see \cite{BZ}. Recall that an
affine isomorphism on ${\mathcal S}(H)$ is a bijective map
$\phi:{\mathcal S}(H) \rightarrow {\mathcal S}(H)$ satisfying
$$\phi(t \rho_1+ (1-t)\rho_2)=t \phi(\rho_1)+(1-t) \phi(\rho_2) \quad \hbox{ for all }
t \in [0,1]  \hbox{ and } \rho_1, \rho_2 \in {\mathcal S}(H).$$
Evidently, we have the implications (c) $\Rightarrow$ (b)
$\Rightarrow$ (a) for a bijective map $\phi: {\mathcal S}(H)
\rightarrow {\mathcal S}(H)$ for the following conditions.

\begin{itemize}
\item[(a)] $\phi([\rho_1, \rho_2]) \subseteq [\phi(\rho_1), \phi(\rho_2)]$
for any $\rho_1, \rho_2 \in {\mathcal S}(H)$.

\item[(b)] $\phi([\rho_1, \rho_2]) =  [\phi(\rho_1), \phi(\rho_2)]$
for any $\rho_1, \rho_2 \in {\mathcal S}(H)$.

\item[(c)] $\phi$ is an affine isomorphism.
\end{itemize}
It was  shown in \cite{BZ} that an affine isomorphism
$\phi:{\mathcal S}(H) \rightarrow {\mathcal S}(H)$ has the form
$$\rho\mapsto U\rho U^* \quad \hbox{ or } \quad   \rho \mapsto U\rho^T U^*,
\eqno{(1.1)}$$ where $U$ is a unitary operator and $\rho^T$ is the
transpose of $\rho$ with respect to a certain orthonormal basis for
$H$. Note that unitary similarity transforms correspond to
evolutions of quantum systems, and many maps that leave invariant
subsets or quantum properties of the states have the form described
in (1.1). One may be tempted to conjecture that maps on states
satisfying (a) or (b) above also have the forms described in (1.1).
However, this is not true as shown by our results. It turns out that
the maps satisfying condition (a) and (b) are closely related to
quantum  measurements.

Recall that in quantum mechanics a fine-grained quantum measurement
is described by a collection  $\{M_m\}$ of measurement operators
acting on the state space $H$ satisfying $\sum_mM_m^*M_m=I$. Let
$M_j$ be a measurement operator. If the state  of the quantum system
is $\rho\in{\mathcal S}(H)$ before the measurement, then the state
after the measurement is $\frac{M_j\rho M_j^*}{{\rm tr}(M_j\rho
M_j^*)}$ whenever $M_j\rho M_j^*\not=0$. If $M_j$ is fixed, we get a
measurement  map $\phi_{j}$ defined by $\phi_j(\rho)=\frac{M_j\rho
M_j^*}{{\rm tr}(M_j\rho M_j^*)}$ from the convex subset ${\mathcal
S}_M(H)=\{ \rho : M_j\rho M_j^*\not=0\}$ of the (convex) set
${\mathcal S}(H)$ of states into ${\mathcal S}(H)$. If $M_j$ is
invertible, then $\phi_{j}:{\mathcal S}(H)\rightarrow{\mathcal
S}(H)$ is bijective and will be called an invertible measurement
map. Observe that a measurement map $\phi_{j}$ satisfies (a), (b),
and is not of the standard form (1.1) in general.

In this paper, we show that, up to the transpose, bijective maps on
states satisfying (a) or (b) are precisely invertible measurement
maps. The following is our main result.

\vspace{5mm}{\bf Theorem 1.} {\it Let ${\mathcal S}(H)$ be the
convex set of all states on Hilbert space $H$ with $2\leq \dim H\leq
\infty$.
 The following statements are equivalent for a bijective   map $\phi:{\mathcal S}(H) \rightarrow {\mathcal S}(H)$.

\begin{itemize}
\item[{\rm (a)}] $\phi([\rho_1, \rho_2]) \subseteq [\phi(\rho_1), \phi(\rho_2)]$
for any $\rho_1, \rho_2 \in {\mathcal S}(H)$.

\item[{\rm (b)}] $\phi([\rho_1, \rho_2]) =  [\phi(\rho_1), \phi(\rho_2)]$
for any $\rho_1, \rho_2 \in {\mathcal S}(H)$.

\item [{\rm (c)}] There is an invertible bounded linear operator $M\in
{\mathcal B}(H)$ such that $\phi$ has the form
$$\rho \mapsto \frac{M\rho M^*}{{\rm tr}(M\rho M )}\quad {\rm or }
\quad \rho \mapsto \frac{M\rho^T M^*}{{\rm tr}( M\rho^TM^*)} ,$$
where $\rho^T$ is the transpose of $\rho$ with respect to an
orthonormal basis.
\end{itemize}
}

It is interesting to note that condition (a) is much weaker than
condition (b). For example, condition (a) does not even ensure that
$\phi([\rho_1, \rho_2])$ is a convex (connected) subset of
$[\phi(\rho_1), \phi(\rho_2)]$. It turns out that the two conditions
(a) and (b) are equivalent for a bijective map, and the map must be
a measurement map or the composition of the transpose map with a
measurement map.

The proof of Theorem 1 is done in the next few sections. In Section
2, we will establish the equivalence of (a) and (b) using a result
of P$\breve{a}$les \cite{Z}. Then we verify the equivalence of (b)
and (c). We treat the finite dimensional case in Section 3. Using
the result in Section 3, we complete the proof for the infinite
dimensional case in Section 4.

\section{The equivalence of the first two conditions}

The implication of (b) $\Rightarrow$ (a) is clear.
We consider the implication (a)$\Rightarrow$(b).

Assume (a) holds.
We will prove that $\phi([\rho,\sigma])= [\phi(\rho),\phi(\sigma)]$
for any quantum states $\rho, \sigma$. If $\rho=\sigma$, it is
trivial. Suppose $\rho\neq\sigma$.

Note that $\rho, \sigma\in {\mathcal S}(H)$  are linearly dependent
if and only if $\rho= \sigma$. So, if $\rho, \sigma$ are linearly
independent, then $\phi(\rho),\phi(\sigma)$ are linearly independent
as $\phi(\rho)\neq\phi(\sigma)$ by the injectivity of $\phi$. Let
${\mathcal{HT}}(H)$ be the real linear space of all self-adjoint
trace-class operators on $H$. As $\phi$ is injective, we must have
$\phi(]\rho,\sigma[)\subset ]\phi(\rho ),\psi(\sigma)[$ for any
$\rho ,\sigma\in{\mathcal S}(H)$, where $]\rho ,\sigma_2[=[\rho
,\sigma]\setminus \{\rho ,\sigma\}$ is the open line segment joining
$\rho ,\sigma$. So by P$\breve{a}$les' result \cite[Theorem 2]{Z},
there exists a real linear map $\psi:{\mathcal
{HT}}(H)\rightarrow{\mathcal {HT}}(H)$, a real linear functional $f:
{\mathcal {HT}}(H)\rightarrow{\mathbb R}$, an operator
$B\in{\mathcal HT}(H)$ and a real number $c$ such that
$$\phi(\rho)=\frac{\psi(\rho)+B}{f(\rho)+c} \quad
\mbox{\rm and}\quad f(\rho)+c>0 \eqno(2.1)$$
hold  for all $\rho\in {\mathcal S}(H)$. Thus, for any $\rho, \sigma \in
{\mathcal S}(H)$ with $\rho\not=\sigma$ and any $t \in [0,1]$, there
exists $s\in [0,1]$ such that
$$\phi(t\rho+(1-t)\sigma)=s\phi(\rho)+(1-s)\phi(\sigma)
=s\frac{\psi(\rho)+B}{f(\rho)+c}+(1-s)\frac{\psi(\sigma)+B}{f(\sigma)+c}.$$
On the other hand, by the linearity of $\psi$ and $f$, we have
\begin{eqnarray*}
\phi(t\rho+(1-t)\sigma)
&=& \frac{\psi(t\rho+(1-t)\sigma)+B}{f(t\rho+(1-t)\sigma)+c} \\
&=& t\frac{\psi(\rho)+B}{f(t\rho+(1-t)\sigma)+c}+
(1-t)\frac{\psi(\sigma)+B}{f(t\rho+(1-t)\sigma)+c}.
\end{eqnarray*}
Write $\lambda_{t, \rho, \sigma}=f(t\rho+(1-t)\sigma)+c$, we get
$$(\frac{s}{f(\rho)+c}-\frac{t}{\lambda_{t, \rho,
\sigma}})(\psi(\rho)+B)+(\frac{1-s}{f(\sigma)+c}-\frac{1-t}{\lambda_{t,
\rho, \sigma}})(\psi(\sigma)+B)=0.$$ As $\rho\neq\sigma$,
$\phi(\rho)$ and $\phi(\sigma)$ are linearly independent. This
implies that $\psi(\rho)+B$ and $\psi(\sigma)+B$ are linearly
independent, too. It follows that
$$\frac{t}{f(t\rho+(1-t)\sigma)+c}=\frac{s}{f(\rho)+c} \ {\rm and }\
\frac{1-t}{f(t\rho+(1-t)\sigma)}=\frac{1-s}{f(\sigma)+c}.$$
Clearly,  $s$ is continuously dependent of $t$ such that
$\lim_{t\rightarrow 0}  s = 0$ and
$\lim_{t\rightarrow 1} s = 1$. Hence we must have
$\phi([\rho,\sigma])=[\phi(\rho),\phi(\sigma)]$. Thus, condition (b)
holds. \qed

Denote by ${\mathcal Pur}(H) = \{x\otimes x: x \in H, \|x\| = 1\}$
the set of pure states in ${\mathcal S}(H)$.
The following lemma is useful for our future discussion.

{\bf Lemma 2.1.} If condition (b) of Theorem 1 holds, then $\phi$ preserves pure states in both
directions, that is, $\phi({\mathcal Pur}(H))={\mathcal Pur}(H)$ .

\bf Proof\  \rm
 It is clear that ${\mathcal S}(H)$ is a convex set and its  extreme point set
 is the set ${\mathcal Pur}(H)$ of all pure states (rank-1 projections).
 For any $P\in{\mathcal Pur}(H)$, if $\phi^{-1}(P)\not\in {\mathcal Pur}(H)$, then there are
 two states $Q, R\in{\mathcal S}(H)$ such that $Q\not=R$ and $\phi^{-1}(P)=tQ+(1-t)R$. As
  $\phi([\rho,\sigma])\subseteq [\phi(\rho),\phi(\sigma)]$ for any $\rho,\sigma$,
 there is some $s\in[0,1]$ such that $P=\phi(\phi^{-1}(P))=\Phi(tQ+(1-t)R)=s\phi(Q)+(1-s)\phi(R)$.
 Since $\phi(Q)\neq \phi(R)$, this  contradicts
  the fact that $P$ is extreme point. So $\phi^{-1}$ sends pure states to pure states.
  Similarly, since $\phi([\rho,\sigma])\supseteq[\phi(\rho),\phi(\sigma)]$
  for any states $\rho, \sigma$,
 one can show   that $\phi$ maps pure states into pure states. \qed

 \section{Proof of Theorem 1: finite dimensional case}

In this section we assume that $\dim H=n<\infty$. In such a case, we
may regard ${\mathcal {HT}}(H)$ the same as ${\bf H}_n$, the real
linear space of $n\times n$ Hermitian matrices. Since the
implication (c) $\Rightarrow$ (b), we needs only prove the
implication (b) $\Rightarrow$ (c). We divide the proof of this
implication into several assertions. Assume (b) holds.

\medskip\noindent
{\bf Assertion 3.1.} $\phi(\frac{I}{n})$ is invertible.

 Let $\phi(\frac{I}{n}) = T$.  In order to prove $T$ is invertible, we show that $\phi$ maps  invertible states to invertible states.
Note that $\phi$ has the form of Eq.(2.1), that is, for any $\rho\in
{\mathcal S}(H)$, $\phi(\rho)=\frac{\psi(\rho)+B}{f(\rho)+c}$. Since
 ${\bf H}_n $ is finite dimensional, the linear map $\psi$ and the
 linear functional $f$ are  bounded. So $\phi$ is
continuous.  $\phi^{-1}$ is also continuous as $\phi$ preserves line
segment and hence has the form of Eq.(2.1). Thus $\phi$ maps open
sets to open sets. Denote by $G({\mathcal S}(H))$ the  subset of all
invertible states. $G({\mathcal S}(H))$ is an open subset of
${\mathcal S}(H)$. In fact, $G({\mathcal S}(H))$ is the maximal open
set of all interior points of ${\mathcal S}(H)$. To see this, assume
that a state $\rho$  is not invertible; then there are mutually
orthogonal rank-one projections $P_i$ ($i=1,2,\ldots n$), an integer
$1\leq k<n$ and scalars $t_i> 0$ with $\sum_{i=1}^kt_i=1$ such that
$\rho=\Sigma^k_{i=1} t_iP_i$.  For any $\varepsilon> 0$ small enough
so that $\frac{\varepsilon}{2k}<\min\{t_1,t_2\ldots ,t_k\}$, let
$$\rho_\varepsilon=\Sigma^k_{i=1}(t_i-\frac{\varepsilon}{2k})P_i+
\Sigma^n_{j=k+1}(\frac{\varepsilon}{2(n-k)})P_j.$$
Then $\rho_\varepsilon$ is an invertible state and
$$\|\rho-\rho_\varepsilon\|_{{\rm tr}}\leq\Sigma^{k}_{i=1}\frac{\varepsilon}{2k}
+\Sigma^n_{j=k+1}\frac{\varepsilon}{2(n-k)}=\varepsilon.$$
It follows that for any  state $\rho$ and any $\varepsilon>0$, there
is an invertible state $\sigma$ such that $\rho\in \{\tau
\in{\mathcal S}(H): \|\tau-\sigma\|_{\rm Tr}< \varepsilon\}$.
 So the trace norm closure of $G({\mathcal S}(H))$ equals  ${\mathcal S}(H)$.
Thus $G({\mathcal S}(H))$ is the set of all interior points of
${\mathcal S}(H)$. Since $\phi$ preserves the open sets, we have
$\phi(G({\mathcal S}(H)))\subseteq G({\mathcal S}(H))$.  So $\phi$
preserves the invertible states. In particular, $\phi(\frac{I}{n}) $
is invertible. \qed

By Assertion 1, there is an invertible operator $R\in{\mathcal B}(H)$ such
that $\phi(\frac{I}{n}) = RR^*$. Let $S = R^{-1}$; then the map
$\tilde \phi :{\mathcal S}(H)\rightarrow{\mathcal S}(H)$ defined by
$$\rho \mapsto \frac{S\phi(\rho)S^*}{{\rm tr}(S\phi(\rho)S^*)}$$ is bijective, sends
line segments to line segments  in both directions, i.e., $\tilde
\phi([\rho,\sigma])= [\tilde \phi(\rho),\tilde \phi(\sigma)]$, and
satisfies $\tilde \phi(\frac{I}{n}) = \frac{I}{n}$.

\medskip\noindent
{\bf Assertion 3.2.} $\tilde \phi$ maps orthogonal rank one
projections to orthogonal rank one  projections.

If $\{P_1, \dots, P_n\}$ is an orthogonal set of rank one
projections satisfying $P_1 + \cdots + P_n = I$, then there are
$t_i\in [0,1]$ $(i=1,...,n)$ with $\Sigma_{i=1}^n t_i=1$ such that
$$ \frac{I}{n}= \tilde \phi(\frac{I}{n}) = \tilde \phi(\frac{(P_1 + \cdots + P_n)}{n})
= t_1\tilde \phi(P_1) + \cdots + t_n\tilde \phi(P_n) \ge t_i \tilde \phi(P_i)$$
for each $i = 1, \dots, n$.
Because $\tilde \phi(P_i)$ is a rank one orthogonal projection
and $I/n - t_i \tilde \phi(P_i)$ is positive semidefinite, we see that
$1/n \ge t_i$ for $i = 1, \dots, n$. Taking trace, we have
$$1 = {\rm tr}(I/n) = \sum_{i=1}^n t_i.$$
Thus, $t_1 = \cdots = t_n = 1/n$. So, $I = \sum_{i=1}^n \tilde \phi(P_i)$.
This implies that $\{\tilde \phi(P_1),\ldots ,\tilde
\phi(P_n)\}$ is an orthogonal set of rank one projections. Hence,
$\tilde \phi$ sends orthogonal rank one  projections to orthogonal
rank one
 projections. \qed

By \cite[Theorem 2]{Z} again, $\tilde \phi$ has the form of
Eq.(2.1), that is,
$$\tilde \phi(\rho)=\frac{\psi(\rho)+B}{f(\rho)+c} \eqno(3.1)$$
holds for any $\rho\in {\mathcal S}(H)$, where $\psi: {\mathbf
H}_n({\mathbb C})\rightarrow {\mathbf H}_n({\mathbb C})$ is a real
linear map, ${\mathbf H}_n({\mathbb C})$ is the real linear space of
all $n\times n$ hermitian matrices, $B\in {\mathbf H}_n({\mathbb
C})$, $f:{\mathbf H}_n({\mathbb C})\rightarrow {\mathbb R}$ is a
real linear functional and $c $ is a real constant with
$f(\rho)+c>0$ for all $\rho\in{\mathcal S}(H)$.

Next we consider the two cases of dim$H >2$ and dim$H=2$
respectively.

\medskip\noindent
{\bf Assertion 3.3.} Assume dim$H> 2$. The functional $f$ in
Eq.(3.1) is a constant on ${\mathcal S}(H)$, that is, there is a
real number $a$ such that $f(\rho)=a$   for all $\rho\in{\mathcal
S}(H)$.

For any normalized orthogonal basis $\{e_i\}^n_{i=1}$, let
$P_i=e_i\otimes e_i$. We first claim that $f(e_i\otimes
e_i)=f(e_j\otimes e_j)$ for any $i$ and $j$. Since $\tilde \phi$
preserves the rank one projections in both directions,
there is a rank one projection $Q_i=x_i\otimes x_i$  such that
$$x_i\otimes x_i=Q_i=\tilde \phi(P_i)=\frac{\psi(e_i\otimes e_i)+B}{f(e_i\otimes e_i)+c}.$$
So $$\psi(e_i\otimes e_i)+B=(f(e_i\otimes e_i)+c)(x_i\otimes x_i).$$
As $\tilde \phi (\frac{I}{n})=\frac{I}{n}$ and
$\frac{I}{n}=\frac{1}{n}\sum^n_{i=1}e_i\otimes e_i$, we have
$$\frac{I}{n}=\tilde \phi (\frac{1}{n}\sum^n_{i=1}e_i\otimes e_i)
=\frac{\psi(\sum^n_{i=1}\frac{1}{n}e_i\otimes
e_i)+B}{f(\sum^n_{i=1}\frac{1}{n}e_i\otimes e_i)+c}
=\frac{\sum^n_{i=1}\frac{1}{n}\psi(e_i\otimes
e_i)+n\frac{1}{n}B}{\sum^n_{i=1}\frac{1}{n}f(e_i\otimes
e_i)+n\frac{1}{n}c}.$$ Then
$$\frac{I}{n}=\frac{\frac{1}{n}(\sum^n_{i=1}\psi(e_i\otimes
e_i)+B)}{\frac{1}{n}(\sum^n_{i=1}f(e_i\otimes
e_i)+c)}=\frac{\sum^n_{i=1}(\psi(e_i\otimes
e_i)+B)}{\sum^n_{i=1}(f(e_i\otimes e_i)+c)}.\eqno(3.2)$$ On the
other hand, by Assertion 3.2, we have
$$\frac{I}{n}=\tilde \phi (\frac{I}{n})=\frac{1}{n}\sum^n_{i=1}\tilde\phi (e_i\otimes e_i)
=\frac{1}{n}\sum^n_{i=1}\frac{\psi(e_i\otimes e_i)+B}{f(e_i\otimes
e_i)+c}.$$ Thus we get
$$I=\sum^n_{i=1}\frac{\psi(e_i\otimes e_i)+B}{f(e_i\otimes
e_i)+c}.\eqno(3.3)$$ Let $A_i=\psi(e_i\otimes e_i)+B$ and
$a_i=f(e_i\otimes e_i)+c$. Then Eq.(3.2) and Eq.(3.3) imply that
$$I=n(\frac{A_1+A_2+\ldots+A_n}{a_1+a_2+\ldots+a_n})=\frac{A_1}{a_1}+\frac{A_2}{a_2}+\ldots+\frac{A_n}{a_n}.$$
Note that $A_i=a_iQ_i$, where $Q_i=\tilde \phi(e_i\otimes e_i)=x_i\otimes x_i$.
Therefore, we get that
$$I= n(\frac{a_1Q_1+a_2Q_2+\ldots+a_nQ_n}{a_1+a_2+\ldots+a_n})
=\frac{a_1Q_1}{a_1}+\frac{a_2Q_2}{a_2}+\ldots+\frac{a_nQ_n}{a_n}.$$
It follows that
$$
n(\frac{a_1Q_1+a_2Q_2+\ldots+a_nQ_n}{a_1+a_2+\ldots+a_n})=Q_1+Q_2+\ldots+Q_n.$$
Since $\{Q_i\}_{i=1}^n$  is an orthogonal set of rank one
projections, we see that
$$\frac{a_1+a_2+\ldots+a_n}{n}=a_1=a_2=\ldots=a_n.$$
This implies that there is some scalar $a$ such that $f(e_i\otimes
e_i)=a$ holds for all $i$. Now for arbitrary unit vectors $x,y\in
H$,  as dim$H>2$, there is a unit vector $z\in H$ such that $z\in
[x,y]^\perp$. It follows from the above argument that $f(x\otimes
x)=f(z\otimes z)=f(y\otimes y)$. So $f(x\otimes x)=a$ for all unit
vectors $x\in H$. Since each state is a convex combination of pure
states, by the linearity of $f$, we get that $f(\rho)=a$ holds for
every state $\rho$. \qed

\medskip\noindent
{\bf Assertion 3.4.} Assume dim$H> 2$. $\phi$ has the form stated in
Theorem 1 (c).

 Every state is a
convex combination of some pure states, i.e.  convex combination of
some rank one projections.  Therefore, by Assertion 3.3, we have
$$\tilde \phi(\rho)=\frac{\psi(\rho)+B}{\alpha+c}$$ holds for all $\rho$.  Then by the linearity of $\psi$, it is clear
that $\tilde\phi$ is an affine isomorphism, i.e., for any states
$\rho, \sigma$ and scalar $\lambda$ with $0\leq\lambda\leq 1$,
$\tilde\phi(\lambda\rho+(1-\lambda )\sigma)=\lambda\tilde
\phi(\rho)+(1-\lambda )\tilde \phi(\sigma).$ By a result due to
Kadison (Ref. \cite[Theorem 8.1]{BZ}), $\tilde \phi$ has the
standard form, that is, there exists a unitary operator
$U\in{\mathcal B}(H)$ such that $\tilde\phi$ has the form
$$\tilde\phi(\rho) = U\rho U^* \ \mbox{\rm for all } \rho\quad \hbox{ or } \quad \rho \mapsto U\rho^TU^*\ \mbox{for all }\ \rho.$$

Now recalled that $\tilde \phi$ is defined by $\tilde \phi(\rho)=
S\phi(\rho)S^*/{\rm tr}(S\phi(\rho)S^*)$. If $\tilde\phi$ takes the
first form, then we have
$$\phi(\rho)= {\rm tr}(S\phi(\rho)S^*)S^{-1}\tilde \phi(\rho)(S^*)^{-1}
= {\rm tr}(S\phi(\rho)S^*)S^{-1}U\rho U^*(S^*)^{-1}.$$
As $1={\rm tr}(\phi(\rho))={\rm tr}(S\phi(\rho)S^*){\rm
tr}(S^{-1}U\rho U^*(S^*)^{-1})$, so $${\rm
tr}(S\phi(\rho)S^*)=\frac{1}{{\rm tr}(S^{-1}U\rho U^*(S^*)^{-1})}.$$
Letting $M=S^{-1}U$, we get $\phi(\rho)=\frac{M\rho M^*}{{\rm
tr}(M\rho M^*)}$ for all $\rho$, that is, $\phi$ has the first form
stated in (c) of Theorem 1.

Similarly, if $\tilde\phi$ takes the second form,
then $\phi$ takes the second form stated in (c) of Theorem 1. \qed

\medskip\noindent
{\bf Assertion 3.5.} Condition (c) of Theorem 1 holds for the case of dim$H=2$.

Assume that  dim$H= 2$. Denote by ${\mathcal S}_2=\mathcal S(H)$ the
convex set of $2\times 2$ positive matrices with the trace 1. Then
the map $\tilde \phi: {\mathcal S}_2\rightarrow {\mathcal S}_2$ is a
bijective map preserving segment in both directions satisfying
$\tilde\phi(\frac{1}{2}I_2)=\frac{1}{2}I_2$. Let us identify
${\mathcal S}_2$ with the unit ball $({\mathbb R}^3)_1 = \{(x,y,z)^T
\in {\mathbb R}^3: x^2+y^2+ z^2 \le 1\}$ of ${\mathbb R}^3$ by the
following way. Let $\pi: ({\mathbb R}^3)_1\rightarrow {\mathcal
S}_2$ be the map defined by
$$(x,y,z)^T\mapsto \frac{1}{2}I_2+\frac{1}{2}\left(\begin{array}{ccccccccccccccc}
z& x-iy\\
x+iy & -z \\
   \end{array}\right).$$
$\pi$ is a bijective  affine isomorphism. Note that $v=(x,y,z)^T$
satisfies $x^2+y^2+z^2 = 1$ if and only if the corresponding matrix
$\pi(v)$ is a rank one projection, and $0=(0,0,0)^T$ if and only if
the corresponding matrix is $\pi(0)=\frac{1}{2}I$. The map $\tilde
\phi: {\mathcal S}_2\rightarrow {\mathcal S}_2$ induces a map
$\hat{\phi}: ({\mathbb R}^3)_1\rightarrow ({\mathbb R}^3)_1$ by the
following equation
$$\tilde\phi(\rho)=\frac{1}{2}I+\pi(\hat{\phi}(\pi^{-1}(\rho))).$$
Since $\tilde\phi$ is a segment preserving bijective map and $\pi$
is an affine isomorphism, the map $\hat\phi$ is a bijective map
preserving segment in both directions, that is,
$\hat\phi([u,v])=[\hat\phi(u), \hat\phi(v)]$ for $u,v\in ({\mathbb
R}^3)_1$. So $\hat\phi$ maps the surface of $({\mathbb R}^{3})_1$
onto the surface of $({\mathbb R}^{3})_1$. Since
$\tilde\phi(\frac{1}{2}I)=\frac{1}{2}I$, we have that
$\hat\phi((0,0,0)^T)=(0,0,0)^T$.

Applying the P$\breve{a}$les' result  \cite[Theorem 2]{Z} to
$\hat\phi$, there exists a linear transformation  $L: {\mathbb
R}^3\rightarrow {\mathbb R}^3$, a linear functional $f:{\mathbb
R}^3\rightarrow {\mathbb R}$, a vector $u_0\in {\mathbb R}^3$ and a
scalar $r\in{\mathbb R} $ such that $f((x,y,z)^T)+r>0$ and
$$\hat \phi((x,y,z)^T)=\frac{L((x,y,z)^T)+u_0}{f((x,y,z)^T)+r}$$
for each $(x,y,z)^T\in ({\mathbb R}^3)_1$.  Since
$\hat\phi((0,0,0)^T)=(0,0,0)^T$, we  have $u_0=0$ and $r>0$.
Furthermore, the linearity of $f$ implies that there are real
scalars $r_1, r_2, r_3$ such that $f((x,y,z)^T)=r_1x+ r_2y+ r_3z$.
We claim that $r_1= r_2= r_3=0$ and hence $f=0$. If not, then there
is a vector $(x_0,y_0,z_0)^T$ satisfying $x^2_0+y^2_0+z^2_0 = 1$
such that $f((x_0,y_0,z_0)^T)=r_1x_0+ r_2y_0+ r_3z_0\neq 0$. It
follows that
$$1=\|\hat
\phi((x_0,y_0,z_0)^T)\|=\|\frac{L((x_0,y_0,z_0)^T)}{r_1x_0+ r_2y_0+
r_3z_0+r}\|,$$
and thus
$$\|L((x_0,y_0,z_0)^T)\|=r_1x_0+ r_2y_0+
r_3z_0+r.$$ Similarly $$\|L((-x_0,-y_0,-z_0)^T)\|=-r_1x_0- r_2y_0-
r_3z_0+r.$$ By the linearity of $L$ we have $r_1x_0+ r_2y_0+
r_3z_0+r=-r_1x_0- r_2y_0- r_3z_0+r$.  Hence $r_1x_0+ r_2y_0+
r_3z_0=0$, a contradiction. So, we have $f=0$, and thus
$\hat\phi=\frac{L}{r}$ is linear. Now it is clear that $\tilde\phi$
is an affine isomorphism as $\pi$ is an affine isomorphism. Applying a
similar argument to the proof of Assertion 3.4 and the Kadison's result,
one sees that $\tilde \phi$
has the standard form.
Thus, Theorem 1 (c) holds. \qed

By Assertions 3.4 and 3.5, we get
the proof of Theorem 1 for finite-dimensional case.

\section{Proof:   infinite dimensional case}

In this section we give a proof of our main result for infinite
dimensional case. Similar to the previous section, we need only establish
the implication (b) $\Rightarrow$ (c). We begin with two lemmas.

\medskip
Let $V_1, V_2$ be linear spaces on a field ${\Bbb F}$, $\upsilon:
{\mathbb F}\rightarrow {\mathbb F}$ a nonzero ring automorphism.
A map $A:V_1\rightarrow V_2$ is called a $\upsilon$-linear operator
if $A(\lambda x)=\upsilon(\lambda)Ax$ for all $x\in V_1$. The
following lemma is similar to \cite[Lemma 2.3.1]{Hou2}.

\vspace{5mm}{\bf Lemma 4.1}  {\it Let $V_1, V_2$ be linear spaces on
a field ${\Bbb F}$, $\tau, \upsilon: {\Bbb F}\rightarrow {\Bbb F}$
nonzero ring auto-isomorphisms. Suppose $A:V_1\rightarrow V_2$ is a
$\tau$-linear transformation, $ B :V_1\rightarrow V_2$ is a
$\upsilon$-linear transformation, and $\dim{\rm span}({\rm ran}(
B))\geq 2$. If $\ker B\subseteq\ker A$ and $Ax$ and $Bx$ are
linearly dependent for all $x\in V$, then $\tau=\upsilon$ and
$A=\lambda B$ for some scalar $\lambda$.}

{\bf Proof} As $\ker B\subseteq\ker A$, for every $x\in V_1$, there
is some scalar $\lambda_x$ such that $Ax=\lambda_x Bx$. If
$Bx\not=0$, then there exists $y\in V_1$ such that $Bx,By$ are
linearly independent. Then $\lambda_{x+y}
(Bx+By)=A(x+y)=\lambda_xBx+\lambda_yBy$. This implies that
$\lambda_x=\lambda_{x+y}=\lambda_y$. Moreover, for any
$\alpha\in{\mathbb F}$, we have $\lambda_{\alpha x}=\lambda_x$. If
$Bx=0$, then $Ax=0$. Thus it follows that there exists a scalar
$\lambda$ such that $Ax=\lambda Bx$ holds for all $x\in V_1$. So,
$A=\lambda B$ and $\tau=\upsilon$.\hfill$\Box$

\vspace{3mm}{\bf Lemma 4.2} {\it Let ${\mathcal S}(H)$ be the set of
all states on Hilbert space $H$ with $\dim H= \infty$, and
$\phi:{\mathcal S}(H) \rightarrow {\mathcal S}(H)$ a bijective map.
If $\phi$ satisfies that, for any $t \in [0,1]$ and $\rho, \sigma
\in {\mathcal S}(H)$, there is $s \in [0, 1]$ such that
$$\phi(t\rho+(1-t)\sigma) = s\phi(\rho) + (1-s)\phi(\sigma),$$ then,
$\phi$ is continuous and there is an invertible bounded linear or
conjugate linear operator $T$ such that
$$\phi(x\otimes x)= \frac{Tx\otimes Tx}{\|Tx\|^2} { \ for\ all\ unit\ vectors\ }x\in H.$$}

{\bf Proof} We complete the proof by checking several assertions.
First we restate Lemma 2.1 as:

\medskip\noindent
{\bf Assertion 4.1.} $\phi$ preserves pure states (rank one
projections) in both directions.

\medskip\noindent
{\bf Assertion 4.2.} For any $x_i\otimes x_i \in{\mathcal Pur}(H)$
with $\{x_1,x_2\ldots ,x_n\}$ linearly independent,  let
$$F(x_1, \ldots , x_n)=C(x_1,\ldots , x_n)\cup F_0(x_1, \ldots , x_n), $$ where $C(x_1,\ldots ,x_n)={\rm cov}\{x_i\otimes
x_i: i=1,2\ldots ,n\}$ is the convex hull of $\{x_i\otimes
x_i\}_{i=1}^n$,
$$\begin{array}{rl} F_0(x_1, \ldots , x_n)= &\{ Z\in {\mathcal
S}(H)\setminus C(x_1,\ldots , x_n)  : \mbox{\rm there exists some }
\\ & W\in{\mathcal S}(H)\setminus C(x_1,\ldots , x_n)\ \mbox{\rm such
that }  [Z,W]\cap C(x_1,\ldots , x_n)\not=\emptyset\}.
\end{array}
$$
 Let $H_0={\rm
span} \{x_1,\ldots ,x_n\}$. Then we have
$$F(x_1, \ldots , x_n)={\mathcal S}(H_0)\oplus
\{0\}. \eqno(4.1)
$$

Obviously, $C(x_1,\ldots ,x_n)\subset {\mathcal S}(H_0)\oplus
\{0\}$. If $Z\in F_0(x_1,\ldots , x_n)$, then there exists some
$W\in{\mathcal S}(H)\setminus C(x_1,\ldots , x_n)$, $t_i>0$ with
$\sum_{i=1}^n t_i=1$ and $t\in (0,1)$ such that
$$ \sum_{i=1}^nt_ix_i\otimes x_i =tZ+(1-t)W.$$ Let $P_0\in  {\mathcal B}(H)$ be the
projection from $H$ onto $H_0$. As $\sum_{i=1}^nt_ix_i\otimes x_i
-tZ=(1-t)W\geq 0$ and $(I-P_0)\sum_{i=1}^nt_ix_i\otimes
x_i=\sum_{i=1}^nt_ix_i\otimes x_i(I-P_0)=0$, we see that
$(I-P_0)Z=Z(I-P_0)=0$, which implies that $P_0ZP_0=Z$ and hence
$Z\in{\mathcal S}(H_0)\oplus \{0\}$.

Conversely, assume that $Z\in{\mathcal S}(H_0)\oplus \{0\}$. Since
$C(x_1,\ldots, x_n)\subset {\mathcal S}(H_0)\oplus \{0\}$, we may
assume that $Z$ is not a convex combination of $\{x_i\otimes
x_i\}_{i=1}^n$. Because $\{x_i\}_{i=1}^n$ is a linearly independent
set, there exists an operator $S\in{\mathcal B}(H_0)$ such that
$\{e_i=Sx_i\}_{i=1}^n$ is an orthonormal basis of $H_0$. Then,
consider
$$S(\sum_{i=1}^na_ix_i\otimes x_i
-Z)S^*=\sum_{i=1}^na_iSx_i\otimes Sx_i
-SZS^*=\sum_{i=1}^na_ie_i\otimes e_i -SZS^*.$$ It is clear that for
sufficient large $a_i>0$, $\sum_{i=1}^na_ie_i\otimes e_i -SZS^* \geq
0$, and hence, $W=\sum_{i=1}^na_ix_i\otimes x_i -Z\geq 0$. This
entails that
$$ \frac{\sum_{i=1}^na_ix_i\otimes x_i}{\sum_{i=1}^n a_i}
=\frac{1}{\sum_{i=1}^n a_i}Z+\frac{{\rm tr}(W)}{\sum_{i=1}^n
a_i}(\frac{W}{{\rm tr}(W)}),
$$
that is, $Z\in F_0(x_1,\ldots , x_n)\subset F(x_1,\ldots ,x_n)$.
This finishes the proof of Eq.(4.1). \qed

\medskip\noindent
{\bf Assertion 4.3.} For any finite-dimensional subspace $H_0\subset
H$, there exists a subspace $H_1$ with $\dim H_1=\dim H_0$ such that
$$\phi({\mathcal S}(H_0)\oplus\{0\})={\mathcal S}(H_1)\oplus \{0\}.$$

Assume that $\dim H_0=n$. Choose an orthonormal basis
$\{x_i\}_{i=1}^n$ of $H_0$. Then by Assertion 4.1, there are unit
vectors $u_i\in H$ such that $\phi(x_i\otimes x_i)=u_i\otimes u_i$.
It is clear that $\{u_i\}_{i=1}^n$ is a linearly independent set.
Let $H_1={\rm span}\{u_i\}_{i=1}^n$. Then $\dim H_1=n$, and by
Eq.(4.1) in Assertion 4.2, we have $F(x_1,\ldots, x_n)={\mathcal
S}(H_0)\oplus\{0\}$, $F(u_1,\ldots, u_n)={\mathcal
S}(H_1)\oplus\{0\}$. Since the bijection $\phi$ preserves segments and
pure states  in both directions, it is easily checked that
$\phi(F(x_1,\ldots, x_n))=F(u_1,\ldots,u_n)$, and the conclusion of
Assertion 4.3 follows. \qed

\medskip\noindent
{\bf Assertion 4.4.} For any finite dimensional subspace $\Lambda
\subset H$, there exists a subspace $H_{\Lambda}\subset H$ with
$\dim H_{\Lambda}=\dim \Lambda$ and an invertible linear or
conjugate linear operator $M_\Lambda:\Lambda\rightarrow H_\Lambda$
such that
$$\phi(P_\Lambda\rho P_\Lambda) =\frac{Q_\Lambda M_\Lambda\rho M_\Lambda^*Q_\Lambda}
{{\rm tr}(M_\Lambda\rho M_\Lambda^*)}$$
for all $\rho\in{\mathcal S}(\Lambda)$, where $P_\Lambda$ and
$Q_\Lambda$ are respectively the projections onto $\Lambda$ and
$H_\Lambda$. Moreover, the $M_\Lambda$ can be chosen so that
$M_{\Lambda_1}=M_{ \Lambda_2}|_{\Lambda_1}$ whenever
$\Lambda_1\subseteq \Lambda_2$.

Let $H_0$ be a finite dimensional subspace of $H$ and let $\{e_1,
e_2, \ldots , e_n\} $ be an orthonormal basis of $H_0$. By Assertion
4.1 there exist unit vectors $ \{u_1,u_2,\ldots , u_n\}$ such that
$\phi(e_i\otimes e_i)=u_i\otimes u_i$. Let $H_1={\rm span}
\{u_1,u_2,\ldots , u_n\}$.
 By Assertion 4.1 again, $\dim
H_1=n=\dim H_0$. It follows from Assertion 4.3 that, for any $\rho
\in{\mathcal S}(H)$, $P_0\rho P_0=\rho$ implies that
$P_1\phi(\rho)P_1=\phi(\rho)$. Thus $\phi$ induces a bijective map
$\phi_0 :{\mathcal S}(H_0)\rightarrow{\mathcal S}(H_1)$ by
$\phi_0(\rho)=\phi(P_0\rho P_0)|_{H_1}$. Applying  Theorem 1 for
finite dimensional case just proved in Section 2, we obtain that
there is an invertible bounded linear operator $M:H_0\rightarrow
H_1$ such that $\phi_0$ has the form
$$\rho \mapsto \frac{M\rho M^*}{{\rm tr}( M^*M\rho)}\quad {\rm or }
\quad \rho \mapsto \frac{M\rho^T M^*}{{\rm tr}(M^*M\rho^T )} ,$$
where $\rho^T$ is the transpose of $\rho$ with respect to the
orthonormal basis $\{e_1, e_2, \ldots , e_n\} $. In the last case,
we let $J: H_0\rightarrow H_0$ be the conjugate linear operator
defined by $J(\sum_{i=1}^n \xi_i e_i)=\sum_{i=1}^n\bar{ \xi_i} e_i$,
and let $M^\prime =MJ$. Then, $M^\prime :H_0\rightarrow H_1$ is
invertible conjugate linear and $\phi_0(\rho)= \frac{M^\prime\rho
{M^\prime}^*}{{\rm tr}( {M^\prime}^*M^\prime\rho)}$ for all
$\rho\in{\mathcal S}(H_0)$. Therefore, the first part of the
Assertion 4.4  is true.

Let $\Lambda_i$, $i=1,2$, are finite dimensional subspaces of $H$
and $M_i $s are associated operators as that obtained above way. If
$\Lambda _1\subseteq \Lambda _2$, then, for any unit vector
$x\in\Lambda_1$, we have $\frac{M_1x\otimes
M_1x}{\|M_1x\|^2}=\phi(x\otimes x)=\frac{M_2x\otimes
M_2x}{\|M_2x\|^2}$. It follows that $M_1x$ and $M_2x$ are linearly
dependent. By Lemma 4.2 we see that $M_2|_{\Lambda_1}=\lambda M_1$
for some scalar $\lambda$. As
$\frac{(\lambda M)\rho (\lambda M)^*}{{\rm tr}( (\lambda M)^*(\lambda M)\rho)}
=\frac{M\rho M^*}{{\rm tr}( M^*M\rho)}$,
we may choose $M_2$ so that $M_2|_{\Lambda_1}= M_1$.
\qed

\medskip\noindent
{\bf Assertion 4.5.} There exists a linear or conjugate linear
bijective transformation $T:H\rightarrow H$ such that
$$\phi(x\otimes x)=\frac{Tx\otimes Tx}{\|Tx\|^2}$$ for every unit
vector $x\in H$ and $T|_\Lambda=M_\Lambda$ for every finite
dimensional subspace $\Lambda$ of $H$.

For any $x\in H$, there is finite dimensional subspace $\Lambda $
such that $x\in\Lambda$. Let $Tx=M_\Lambda x$. Then, by Assertion
4.4, $T:H\rightarrow H$ is well defined, linear or conjugate linear.
And
 by Assertion 4.1, $T$ is bijective.

Note that P$\breve{a}$les' result (Theorem 2 in \cite{Z}) holds true
for the infinite dimensional case. Since $\phi$ preserves segment,
by \cite[Theorems 1-2]{Z}, there exists a linear operator $\Gamma:
{\mathcal {HT}}(H)\rightarrow {\mathcal {HT}}(H)$, a linear
functional $g: {\mathcal {HT}} (H)\rightarrow{\mathbb R}$, a scalar
$b\in{\mathbb R}$ and some operator $B\in\mathcal {HT}(B)$ such that
$$\phi(\rho)=\frac{\Gamma\rho+B}{g(\rho)+b}\eqno(4.2)$$
for all $\rho\in{\mathcal S}(H)$, where
 ${\mathcal {HT}}(H)$ denotes the set of all self-adjoint Trace-class operators
 in ${\mathcal B}(H)$ and $g(\rho)+b>0$ for all $\rho\in{\mathcal S}(H)$.
\qed

\medskip\noindent
 {\bf Assertion 4.6.} The functions $g$, $\Gamma$ in Eq.(4.2) are bounded and hence
$\phi$ is continuous.

 Note that, for any $\rho_1,\rho_2\in{\mathcal S}(H)$ and any $t\in(0,1)$, there exists some
$s(t)\in(0,1)$ such that
$$\phi(t\rho_{1}+(1-t)\rho_{2})=s(t)\phi(\rho_{1})+(1-s(t))\phi(\rho_2).$$
Combining this with Eq.(4.2), one gets
$$\frac{t \Gamma \rho_{1}+(1-t) \Gamma \rho_{2}+B}{t g(\rho_{1})+(1-t)
g(\rho_{2})+b}=s(t)\frac{\Gamma\rho_{1}+B}{g(\rho_{1})+b}+(1-s(t))\frac{\Gamma\rho_{2}+B}{g(\rho_{2})+b}.\eqno(4.3)$$
Note that different states are linearly independent. Comparing the
coefficients of $\Gamma\rho_{1}$ in Eq.(4.3), one  sees that
$$s(t)=\frac{t(g(\rho_{1})+b)}{tg(\rho_{1})+(1-t)g(\rho_{2})+b}.\eqno(4.4)$$
It follows that $s(t)\rightarrow 1$ when $t\rightarrow 1$. If
$\rho=\sum_{i=1}^{n}t_{i}\rho_{i}\in{\mathcal S}(H)$ with $\rho_i\in
{\mathcal S}(H)$, one can get some $p_i$ so that
$\phi(\rho)=\phi(\sum_{i=1}^{n}t_{i}\rho_{i})=\sum_{i=1}^{n}p_{i}\phi(\rho_{i})$,
where $\sum_{i=1}^{n}t_{i}=\sum_{i=1}^{n}p_{i}=1$. Similarly we can
check that
$$p_{i}=\frac{t_{i}(g(\rho_{i})+b)}{\sum_{i=1}^{n}t_{i}g(\rho_{i})+b}.\eqno(4.5)$$

Suppose that  $\rho, \rho_i\in{\mathcal S}(H)$ with
$\rho=\sum_{i=1}^{\infty}t_{i}\rho_{i}$, where $t_i>0$ and
$\sum_{i=1}^{\infty}t_{i}=1$. Then
$$\begin{array}{rl}\phi(\rho)=&\phi(\sum_{i=1}^{\infty}t_{i}\rho_{i})\\
=&
\phi((\sum_{j=1}^{k}t_{j})\sum_{i=1}^{k}(\frac{t_{i}}{\sum_{j=1}^{k}t_{j}})
\rho_{i}+(1-\sum_{j=1}^{k}t_{j})\sum_{i=k+1}^{\infty}(\frac{t_{i}}{1-\sum_{j=1}^{k}t_{j}})\rho_{i})\\
=&s_{k}\phi(\sum_{i=1}^{k}(\frac{t_{i}}{\sum_{j=1}^{k}t_{j}})\rho_{i})+(1-s_{k})\phi(\sum_{i=k+1}^{\infty}
(\frac{t_{i}}{1-\sum_{j=1}^{k}t_{j}})\rho_{i}).
\end{array}\eqno(4.6)$$ Thus there exist scalars $q_i^{(k)}>0$
with $\sum_{i=1}^{k} q_i^{(k)}=1$ such that
$$s_{k}\phi(\sum_{i=1}^{k}(\frac{t_{i}}{\sum_{j=1}^{k}t_{j}})\rho_{i})=\sum_{i=1}^{k}s_{k}q_{i}^{(k)}\phi(\rho_{i}).$$
According to Eq.(4.4), Eq.(4.5), and keeping in mind that $g$ is a
linear functional,  a simple calculation reveals that
$$\begin{array}{rl}
 s_{k}=&\frac{(\sum_{j=1}^{k}t_{j})(g(\sum_{i=1}^{k}(\frac{t_{i}}{\sum_{j=1}^{k}t_{j}})\rho_{i})+b)}
{(\sum_{j=1}^{k}t_{j})(g(\sum_{i=1}^{k}(\frac{t_{i}}
{\sum_{j=1}^{k}t_{j}})\rho_{i}))+(1-\sum_{j=1}^{k}t_{j})
g(\sum_{i=k+1}^{\infty}(\frac{t_{i}}{1-\sum_{j=1}^{k}t_{j}})\rho_{i}))+b}\\
=&
\frac{(\sum_{j=1}^{k}t_{j})(g(\sum_{i=1}^{k}(\frac{t_{i}}{\sum_{j=1}^{k}t_{j}})\rho_{i})+b)}
{g(\rho)+b},
\end{array}
 \eqno(4.7)
$$
$$q_i^{(k)}=\frac{(\frac{t_{i}}{\sum_{j=1}^{k}t_{j}})(g(\rho_{i})+b)}
{\sum_{i=1}^{k}(\frac{t_{i}}{\sum_{j=1}^{k}t_{j}})g(\rho_{i})+b}=
\frac{(\frac{t_{i}}{\sum_{j=1}^{k}t_{j}})(g(\rho_{i})+b)}{g(\sum_{i=1}^{k}(\frac{t_{i}}{\sum_{j=1}^{k}t_{j}})\rho_{i})+b},\eqno(4.8)$$
and
$$s_{k}q_{i}^{(k)}=\frac{t_{i}(g(\rho_{i})+b)}{g(\rho)+b}.\eqno(4.9)$$
Observe that $s_{k}q_{i}^{(k)}$ is independent to $k$. Since
$\sum_{i=1}^{k}t_{i}\rightarrow 1$ as $k\rightarrow \infty$, we
must have $s_{k}\rightarrow 1$ as $k\rightarrow \infty$.
Eqs.(4.6)-(4.9) imply that
$$\sum_{i=1}^{\infty}\frac{t_{i}(g(\rho_{i})+b)}{g(\rho)+b}=1$$ and
$$\phi(\sum_{i=1}^\infty t_i\rho_i)=\sum_{i=1}^{\infty}(\frac{t_{i}(g(\rho_{i})+b)}{g(\rho)+b})\phi(\rho_i).\eqno(4.10)$$
In particular, we have $$g(\sum_{i=1}^\infty
t_i\rho_i)=\sum_{i=1}^\infty t_ig(\rho_i).\eqno(4.11)$$

We assert that $\sup \{g(\rho): \rho\in{\mathcal S}(H)\}<\infty$.
Assume  that  $\sup \{g(\rho): \rho\in{\mathcal
S}(H)\}=\infty$. Then, for any positive integer $i$, there exists
$\rho_{i}\in {\mathcal S}(H)$ satisfying that $g(\rho_{i})>2^i$. Let
$\rho_0=\sum_{i=1}^{\infty}\frac{1}{2^{i}}\rho_{i}$,
$\sigma_{k}=\sum_{i=1}^{k}\frac{1}{2^{i}}\rho_{i}$, then
$\sigma_{k}\rightarrow\rho_0$, and
$$g(\sigma_{k})=\sum_{i=1}^{k}\frac{1}{2^{i}}g(\rho_{i})\geq
\sum_{i=1}^{k} 1=k. $$ Since $g(\rho_{i})\geq 0$,  by Eq.(4.11), we
have $g(\rho_0)\geq g(\sigma _k)\geq k$ for every $k$, contradicting
to the fact that $g(\rho_0)<\infty$.  Now the fact $g(\rho)+b>0$ for
all $\rho$ entails that there exists a positive number $c$ such that
$\sup \{|g(\rho)|: \rho\in{\mathcal S}(H)\}=c$. Thus $g$ is
continuous on $  {\mathcal {HT}}(H)$ and
$$\|g\|=c<\infty.\eqno(4.12)$$
Since $$ \|\Gamma \rho\|\leq \|\Gamma \rho +B\|+\|B\|\leq\|\Gamma
\rho + B\|_{\rm Tr}+\|B\|=g(\rho)+b+\|B\|\leq c+|b|+\|B\|$$ holds
for all $\rho\in{\mathcal S}(H)$,  it follows that $\Gamma$ is
$\|\cdot\|_{\rm tr}$-$\|\cdot\|$ continuous from $  {\mathcal
{HT}}(H)$ into itself. Hence, if $\rho_n,\rho\in{\mathcal S}(H)$ and
$\|\cdot\|_{\rm tr}$-$\lim _{n\rightarrow\infty}\rho_n=\rho$, then
$\|\cdot\|$-$\lim_{n\rightarrow\infty}\phi(\rho_n)=\phi(\rho)$.
However, convergence under trace-norm topology and convergence under
uniform-norm topology are the same for states \cite{ZM}. Hence we
have $\|\cdot\|_{\rm
tr}$-$\lim_{n\rightarrow\infty}\phi(\rho_n)=\phi(\rho)$, i.e.,
$\phi$ is continuous under the trace-norm topology. \qed

\medskip\noindent
{\bf Assertion 4.7.} The operator $T$ in Assertion 4.5 is bounded.

For any finite dimensional subspace $\Lambda \subset H$, let
$M_\Lambda$ be the invertible linear or conjugate linear operator
stated in Assertion 4.4. Then for any $\rho\in{\mathcal S}(H)$ with
range in $\Lambda$, we have $\frac{\Gamma \rho
+B}{g(\rho)+b}=\frac{Q_\Lambda M_\Lambda \rho
M_\Lambda^*Q_\Lambda}{{\rm tr}(M_\Lambda \rho M_\Lambda^*)}$. Thus
$$\Gamma \rho +B=\lambda _\rho Q_\Lambda M_\Lambda \rho
M_\Lambda^*Q_\Lambda,$$ where
$$\lambda_\rho=\frac{g(\rho)+b}{{\rm tr}(M_\Lambda \rho
M_\Lambda^*)}.$$ For any $\sigma\in{\mathcal S}(H)$ with range in
$\Lambda$ and $\sigma\not= \rho$, and for any $0<t<1$, by
considering $t\rho+(1-t)\sigma$ one gets
$$\lambda _\rho=\lambda_{t\rho+(1-t)\sigma}=\lambda_\sigma.$$ This
implies that there exists a scalar $d>0$ such that $\lambda _\rho=d$
for all $\rho$ with range in $\Lambda$. Use Assertion 4.4 again, it
is clear that $d$ is not dependent to $\Lambda$. Thus, the equation
$$ {\rm tr}(M_\Lambda \rho M_\Lambda^*)=d^{-1}(g(\rho)+b)$$
holds for all finite rank $\rho\in {\mathcal S}(H)$.
In particular, for any unit vector $x\in \Lambda$,
by Assertion 4.6, $\|g\|<\infty$ and we have
$$ \|M_\Lambda x\|^2=d^{-1}(g(x\otimes x)+b)\leq d^{-1}(\|g\|+|b|)<\infty,$$
which implies that $\|M_\Lambda\|\leq
\sqrt{d^{-1}(\|g\|+|b|)}.$ It follows that, for any unit vector
$x\in H$, we have $\|Tx\|\leq \sqrt{d^{-1}(\|g\|+|b|)}$ and hence
$\|T\|\leq \sqrt{d^{-1}(\|g\|+|b|)}$.

The proof is finished.\qed

Now we are in a position to give a proof of the main theorem for
infinite dimensional case.

{\bf Proof of Theorem 1: infinite dimensional case.}  Similar to the
finite dimensional case, we need only to show (b) $\Rightarrow$ (c).

Assume (b). By Lemma 4.2, there is a bounded invertible linear or
conjugate linear operator $T$ such that $\phi(x\otimes x)=
\frac{Tx\otimes Tx}{\|Tx\|^2}= \frac{Tx\otimes xT^*}{\|Tx\|^2}$  for
all unit vectors $x\in H$. Let $\rho$ be any finite rank state. Then
there exists a finite dimensional subspace $\Lambda $ of $H$ such
that the range of $\rho$ is contained in $\Lambda$. By Assertion 4.4
in the proof of Lemma 4.3, we have $\phi(\rho)=\frac{(Q_\Lambda
M_\Lambda)\rho (Q_\Lambda M_\Lambda)^*}{{\rm tr}((Q_\Lambda
M_\Lambda)\rho (Q_\Lambda M_\Lambda)^*)}=\frac{T\rho T^*}{{\rm
tr}(T\rho T^*)}$. Since the set of finite-rank states is dense in
${\mathcal S}(H)$ and, by Lemma 4.3, $\phi$ is continuous, we get
that $\phi(\rho)=\frac{T\rho T^*}{{\rm tr}(T^*T\rho )}$ for all
states $\rho$ as desired, completing the proof.\qed

\smallskip

\medskip\noindent
{\bf Acknowldegement}

We thank Wen-ling Huang for helpful discussions on the segment preserving
bijections $\hat{\phi}$ in Assertion 3.5.


\smallskip
\begin{thebibliography}{www}


\bibitem{A} E. Alfsen, F. Shultz, Unique decompositions, faces, and
automorphisms of separable states, Journal of Mathematical Physics,
51(2010), 052201.

\bibitem{BZ} I. Bengtsson, K. Zyczkowski, Geometry of quantum states:
an introduction to quantum entanglement, Cambridge University Press,
Cambridge, 2006.

\bibitem{Fa} C.-A. Faure, An elementary proof of the fundamental theorem of projective geometry, Geom.
Dedicata, 90(2002), 145-151.

\bibitem{F} S. Friedland, C.-K. Li Y.-T. Poon and N.-S. Sze, The automorphism
group of separable states in quantum information theory, Journal of
Mathematical Physics,  52(2011), 042203.


\bibitem{G} S. Gudder, A structure for quantum measurements, Reports on Mathematical Physics, 55(2005) 2, 249-267.

\bibitem{Hou} J. C. Hou, A characterization of positive linear maps and criteria of entanglement for quantum states,
J. Phys. A: Math. Theor., 43(2010) 385201.

\bibitem{Hou2} J. C. Hou, J. L. Cui, Introduction to linear maps on operator algebras, Scince Press in China, Beijing, 2004.

\bibitem{m} L. Moln$\acute{a}$r,  Characterizations of the automorphisms of Hilbert space effect algebras,
Commun. Math. Phys., 223(2001), 437-450.

\bibitem{m1} L. Moln$\acute{a}$r, On some automorphisms of the set of effects on Hilbert space. Lett. Math.
Phys., 51(2000), 37-45

\bibitem{m2} L. Moln$\acute{a}$r, W. Timmermann, Mixture preserving maps on von Neumann algebra effects, Lett. Math.
Phys., 79(2007), 295-302

\bibitem{NC}
M. A. Nielsen and I. L. Chuang, Quantum Computation and Quantum
Information, Cambridge University Press, 2000.

\bibitem{Z} Z. P$\breve{a}$les,  Characterization of segment and convexity preserving maps,
preprint.

\bibitem{Uh} U. Uhlhorn, Representation of symmetry transformations in quantum
mechanics, Ark. Fysik, 23(1963), 307-340.

\bibitem{W} E. P. Wigner, Group theory: And its application to the quantum
mechanics of atomic spectra, Academic Press, 1959.

\bibitem{ZM}   S. Zhu, Z.-H. Ma, Topologies on quantum states, Phys. Lett. A, 374(2010), 1336-1341.


\end{thebibliography}
\end{document}